\documentclass[runningheads]{llncs}
\usepackage[T1]{fontenc}
\usepackage{graphicx}
\begin{document}
\title{Local News Hijacking: 
\\A Review of International Instances}
\titlerunning{Local News Hijacking}
%
\author{Christine Sowa Lepird\inst{1}\orcidID{0000-0002-9350-5627} \and
Kathleen M. Carley\inst{1}\orcidID{0000-0002-6356-0238}}
%
\authorrunning{C. Lepird et al.}

%
\institute{Carnegie Mellon University, Pittsburgh PA 15213, USA 
\url{https://www.casos.cs.cmu.edu/}}

\maketitle              
\begin{abstract}
In the rise of the digital era, it's easier than ever to create nefarious websites to spread misinformation. A more recent phenomenon in the United States has been the creation of inauthentic \textit{local} news websites to further an information operation campaign. This paper is a review of the 7 instances in which local news websites were created to influence residents of a region between 2007 and 2024. By breaking down the ways in which these sites operated, we discovered commonalities in the approach - resurrecting ``zombie" papers that were previously established authentic local news organizations, sharing these sites on social media, and using website templates from WordPress. By analyzing these commonalities, we propose ways to mitigate the occurrence of these campaigns in the future. 

\end{abstract}
\section{Introduction}

Since 2019, there has been an increased awareness in the United States of partisan organizations spreading ``local" news to swing states; a worrying trend since local news is the most trusted news source for Americans \cite{gottfried_partisan_2021}. To date, over a thousand of these websites that specialize in automated, low-quality reporting, have been discovered \cite{bengani_hundreds_2019}. Reporter Ryan Smith dubbed these sites as ``pink slime"  \cite{tarkov_journatic_2012} invading the American news diet. In addition to their websites, pink slime  has been shared on Facebook, Reddit, and Twitter/X \cite{lepird2024midterms}. 

While this phenomena is new in the United States, the concept of infiltrating local news websites to share political propaganda is not new nor limited to the USA. In this paper, we compile all other known cases of this untoward form of local journalism as an information operation campaign. By understanding the ways in which bad actors are able to infiltrate the local news ecosystem, policy leaders can create rules and guidelines to prevent local news from being hijacked in the future.

\section{Methodology}

In order to find examples of pink slime abroad, we chose to focus on keywords other than ``pink slime" since it is an American-ized term. Instead, we searched for terms that captured the essence of the creation of an information operation campaign going after trusted local news sources with the phrases: ``fake local news", ``hijacked local news" and ``infiltrating local news". These phrases were searched on Google Scholar, Google News, and the social media platform X; it was important to expand beyond academic publications, as many of the articles about these campaigns were done by fellow news reporters. Our selection criterion was to find instances of deceitful local news websites by non-local reporters that was trying to influence a specific community. Overall 7 such instances were found.

\section{International Instances of Pink Slime}

The map below in Figure \ref{map} shows all of the actors and regions involved in the seven campaigns discovered. In this section, we describe each campaign in greater detail in chronological order.
 
\begin{figure}
\includegraphics[width=\textwidth]{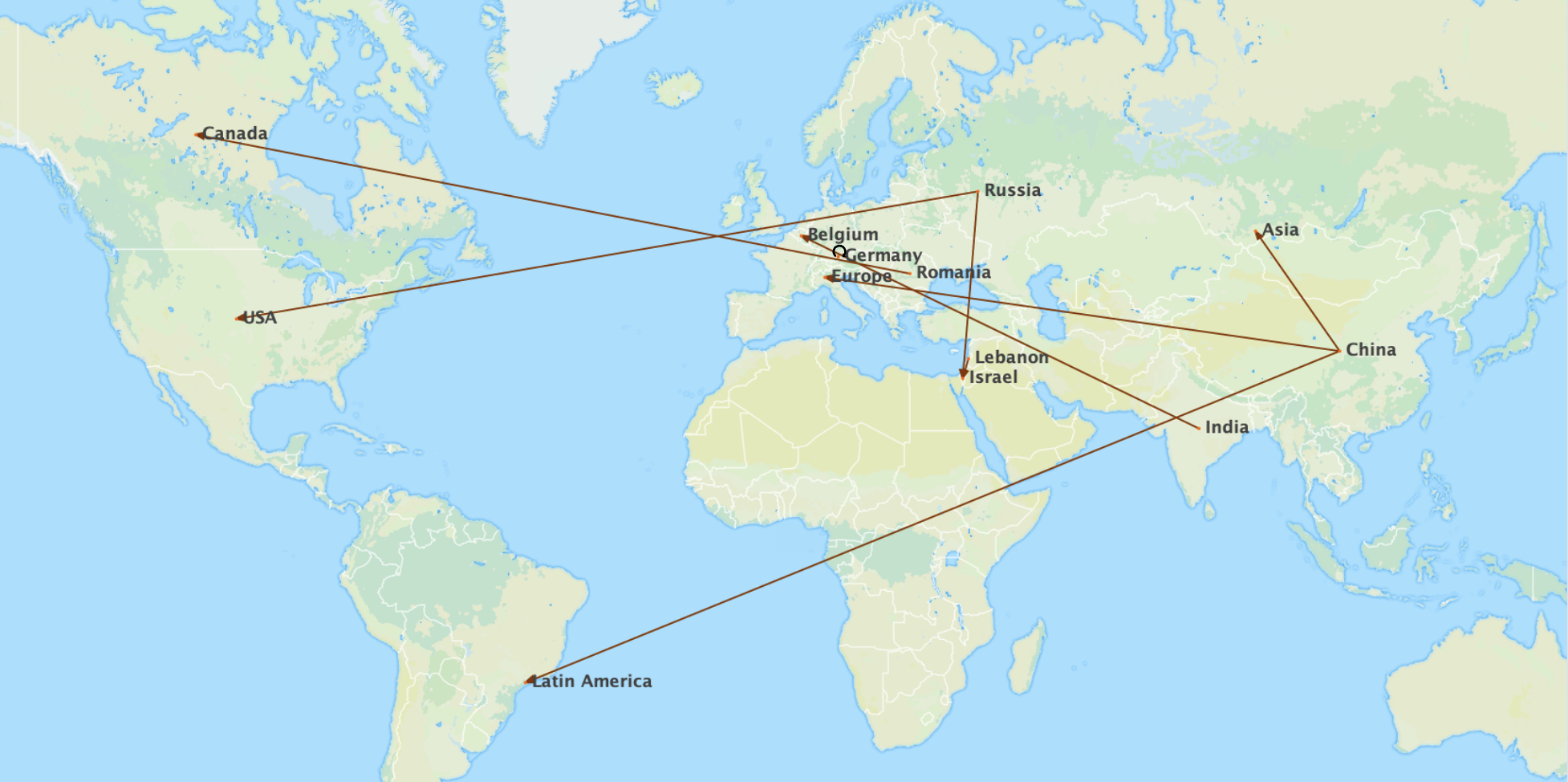}
\caption{A map representing the sources of inauthentic local news and the regions where they created the ``local" news. The geographic regions the arrows point towards are the ones that were infiltrated by the source nodes. Image generated using the ORA network visualization software \cite{carley_ora_2018}.} 
\label{map}
\end{figure}

\subsection{2007: Germany's `Zombie' Papers}
The earliest known example of inauthentic local news dates back to 2007 in Germany. Unlike our other examples, this one has a less-sinister and, unfortunately, more practical origin story. Due to financial pressures, local newsrooms began laying off their reporters, writers, and staff \cite{Assmann}. Eventually, many of these newsrooms had no physical presence in a region but maintained their websites which largely contained stories copied by competitors and money-generating ads \cite{Assmann}. Citizens who were aware of these ``zombie" newsrooms grew distrustful of local news, a painful consequence \cite{Assmann}. Figure \ref{german-zombie-2012} shows what one of these sites looked like before it was taken over by non-local reporters and Figure \ref{german-zombie-2016} shows what the same news site looks like a few years after the takeover. These sites still exist throughout Germany and may have served as a template for later cases of international pink slime, showing organizations that local reporters and a physical presence are not required to run a local news agency \cite{Assmann}. 

\subsection{2010: A Pro-India Campaign in the European Union}

In 2019, the European Union Disinfo Lab uncovered an Indian campaign to influence the European Union by creating 265 ``local" news sites within 65 countries, dating back to 2010 \cite{eudisinfo20191213}. The websites (most of which were named after extinct, real local newspapers, the German ``zombie" approach) shared anti-Pakistan content on their websites as well as associated Twitter accounts \cite{eudisinfo20191213}, as seen in Figures \ref{eptoday} and \ref{eptoday_about}.  

\subsection{2018: Romania Driving Canadian Misinformation}

Initially, Canadians believed that there was a new local news site about updated driver laws in the country \cite{Cadier_2020}; however, the scheme unearthed was much more sinister and involved actors from Romania creating twelve ``local" news websites in Canada, as shown in Figure \ref{canada-eh} \cite{Silverman_2019}. These websites used WordPress templates to share misinformation about recalls, immigration, and driving laws \cite{Cadier_2020} \cite{Silverman_2019}. The articles garnered much interaction on social media platforms like Facebook \cite{Cadier_2020}. While the sites did not venture into promoting political agendas, those researching the phenomena believe that was a next step \cite{Silverman_2019}.

\subsection{2019: The Great Chinese Paperwall Against the World}

A 2024 discovery by The Citizen Lab in Toronto found a pro-Beijing campaign of 123 ``local" news websites in 30 countries operated by a PR Firm in China \cite{Fittarelli2024}. These sites were largely created using WordPress and contained both local and national news stories aggregated from other news sources so as to not draw suspicion with its own original content which contained targeted attacks and conspiracy theories \cite{Fittarelli2024}. An example of one of their sites targeting residents of Venice, Italy and the information it shared about Chinese President Xi Jiping can be found in Figures \ref{great_firewall} and \ref{great_firewall_xijinping}.



\subsection{2023: Russia Infiltrates Israel}

During the Russian-Ukraine conflict, the Russian government worked to change the narrative by creating three ``local" news websites in Israel \cite{Kahan_2023}. These news sites mimic more well known Israeli news sites but include anti-Ukrainian propaganda \cite{Kahan_2023}. Furthermore, the Russians websites ran fake stories accusing U.S. President Biden of trying to ``topple the Israeli government" \cite{Kahan_2023}, further attempting to create a wedge between Israel and the United States.

\subsection{2024: Lebanon \textit{Also} Infiltrates Israel}

As tension in Israel heated up, so did the influence campaigns. Lebanon created ``Dofrek TV", a website designed to share news with Israelis, as seen in Figure \ref{lebanon-israel} \cite{Heller_2024}. In only a few days, news from the site was shared on many social media platforms \cite{Heller_2024}. Despite claiming to be a voice for Israelis, the messaging is anti-Israel, heavily critical of Prime Minister Netanyahu, and pro-Palestinian \cite{Heller_2024}. The majority of the news content is lifted directly from other Israeli news outlets \cite{Heller_2024}. 


\subsection{2024: Russia Comes for America}
In the most recent instance, Russia created 4 news sites that appeared to be local news for 4 major U.S. cities - D.C., New York, Chicago, and Miami - in an attempt to influence the upcoming 2024 U.S. Presidential Election
\cite{Myers_2024}. These sites used WordPress templates, and the Chicago site was akin to another instance of a ``zombie" paper as the Chicago Chronicle was a reputable local newspaper from 1895-1907 (sadly too early to register a news domain on the World Wide Web) that shuttered due to low profits before becoming a ploy in Russian propaganda \cite{Myers_2024}. The group who discovered these sites wisely surmised that ``The purpose is not to fool a discerning reader into diving deeper into the website, let alone subscribing. The goal instead is to lend an aura of credibility to posts on social media spreading the disinformation"\cite{Myers_2024}, a goal the group accomplished. While many of the articles were lifted from other national news sources, some of the original content still included ChatGPT prompts within the text. An example of one of their sites can be seen in Figure \ref{dc-weekly}.

\section{Commonalities}

\begin{figure}
\includegraphics[width=\textwidth]{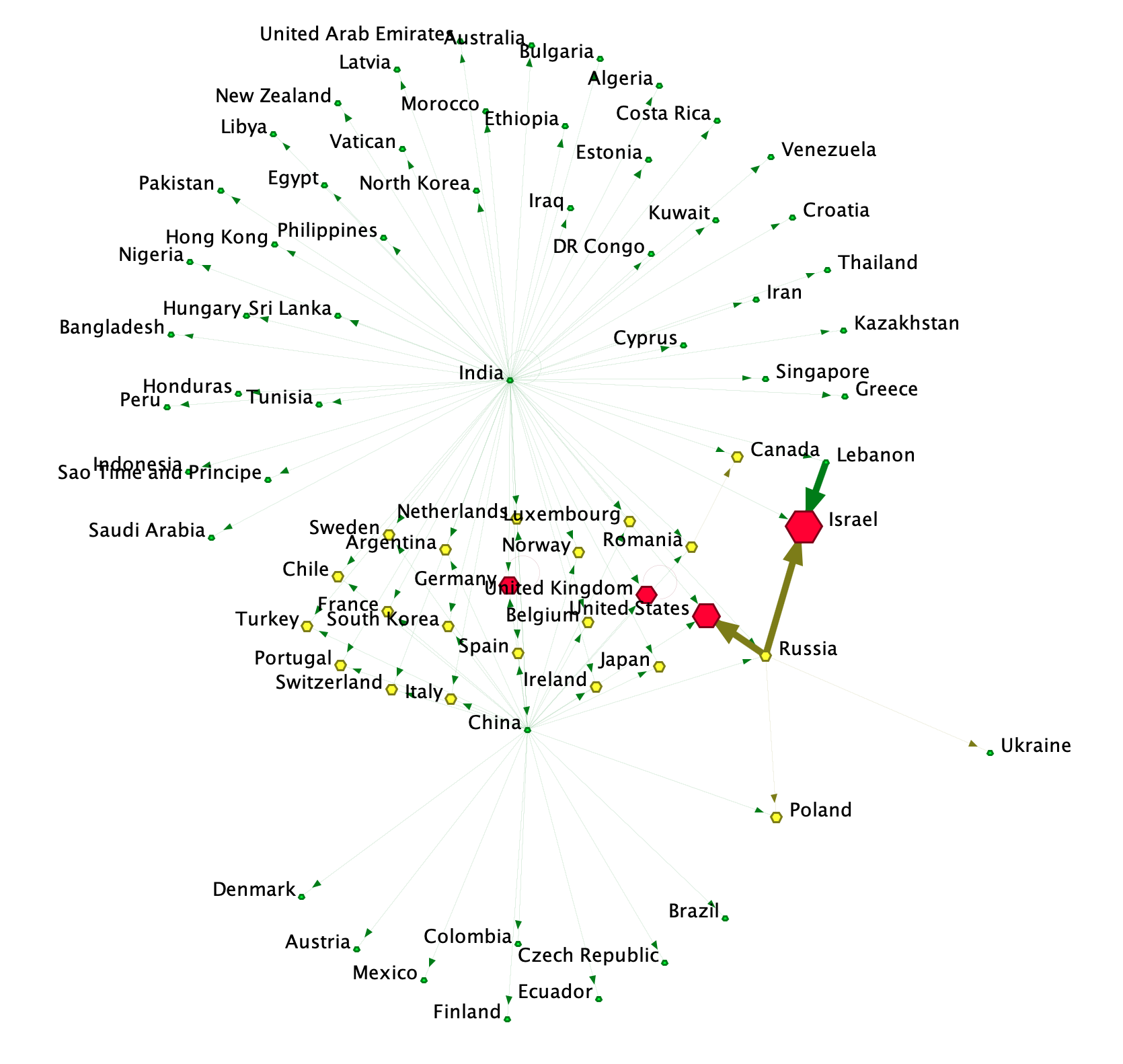}
\caption{A network visualizing all of the individual countries creating and being attacked by local news hijacking campaigns. Red nodes were attacked at least 3 times, yellow nodes were attacked 2 times, and the green nodes were only attacked 1 time. Image generated using the ORA network visualization software \cite{carley_ora_2018}.} 
\label{intl-network}
\end{figure}

The map in Figure \ref{map} shows all of the creating actors and regions attacked in the seven campaigns discovered. Many of these campaigns, like the ones from China and India, targeted dozens of countries, so a more detailed network was created to illustrate all of the countries who fell victim to fake local news campaigns. The network in Figure \ref{intl-network} shows all of the creating actors and individual countries attacked in the seven campaigns discovered. 

To understand the differences between those countries creating fake local news and those who were victims of these attacks, I analyzed the differences in these countries' freedom of the press, democracy, and cyber-security. For freedom of the press, I used the Reporters Without Borders World Press Freedom Index 2024 who bases the index upon ``a score ranging from 0 to 100 that is assigned to each country or territory, with 100 being the best possible score (the highest possible level of press freedom) and 0 the worst" \cite{RWB}. To measure democracy, an index by the Economist Intelligence Unit (2006-2023) was used; it scores countries based on their ability to fairly elect their political leaders and enjoy civil liberties, ranging from 0 to 10 (most democratic) \cite{democracy-index}. Finally, as a measure of cyber-security, the Cybersecurity Exposure Index (CEI) was analyzed to compare the countries' exposure to cyber crime; this index ranges from 0 to 1 (higher exposure) \cite{kateryna_meleshenko_2023}. The results can be found in Table \ref{intl-network}, with a final column added to specifically analyze those countries who have been repeatedly targeted in these attacks. Overall, countries creating fake local news websites enjoy less freedom of the press and democracy and are more subject to cyber crime than their victims. Furthermore, those countries that have been victims of multiple fake local news campaigns enjoy the highest freedom of the press and democracy and lowest exposure to cyber crime. This may suggest that countries with the most freedom of the press and ability to criticize their democracies are more prone to attacks of harmful actors claiming to be ``press" to enjoy the many liberties that those countries afford to such members. 

\begin{table}
\centering
\begin{tabular}{|l|c|c|c|}
\hline
 &
  \textbf{\begin{tabular}[c]{@{}c@{}}Fake Local News \\ Creating Countries\end{tabular}} &
  \textbf{\begin{tabular}[c]{@{}c@{}}All Fake Local News \\ Victim Countries\end{tabular}} &
  \textbf{\begin{tabular}[c]{@{}c@{}}Repeat Fake Local News \\ Victim Countries\end{tabular}} \\ \hline
\textbf{Press Freedom Index}          & 46.5  & 58.2  & 72.4  \\ \hline
\textbf{Democracy Index}              & 5.06  & 6.38  & 8.06  \\ \hline
\textbf{\begin{tabular}[c]{@{}l@{}}Cybersecurity \\ Exposure Index\end{tabular}} & 0.481 & 0.412 & 0.262 \\ \hline
\end{tabular}
\caption{Average index values for creators and victims of fake local news attacks. }
\label{tab:intl-stats}
\end{table}

\section{Recommendations}

Upon reviewing the seven examples of international local news hijacking, we see several commonalities emerge which should be considered for policy action. 

First, the rise of zombie papers. As termed by \cite{Assmann} to describe the happenings in Germany, we saw other examples of foreign groups invading local news markets by using the names, logos, or likenesses of previously-active local news sites for the region. The European DisinfoLab suggested that we ``urge the domain name industry to seriously reflect on this kind of fraudulent, disinforming behaviour as technical abuse of the domain name system" \cite{eudisinfo2}.  The United States is in a unique position to counter these zombies through the use of legislation like the Anticybersquatting Consumer Protection Act \cite{americans_1999}. While the law was passed in 1999, prior to the creation of these sites, it makes the action of creating a domain name that is in violation of a trademark illegal; however, the law fails to consider the international scope of these crimes. By passing legislation that classifies these international ``local" news sites as cyber warfare, they can be removed from the online news ecosystem swiftly. 

Second, these sites are all shared on social media. Many social media powerhouses have relied on Section 230 of the 1996 Communications Decency Act to protect them from being tried as a publisher of what their users share on their platforms. However, Facebook, Twitter, and Reddit have taken action against previous information operation campaigns that they are made aware of. Having a ``tip" line for these instances that is shared with all of the social media platforms' content moderation teams could expedite the removal of their content or at the least, flagging it (or adding a ``Community Note" in the case of Twitter/X). During the Covid-19 pandemic in 2020, Facebook announced that they would be providing ratings of ``Altered", ``Missing Context", ``False", and ``Partly False"  via fact-checking partners to counter misinformation on their site, indicating that further labeling of pink slime would not be out of line with their news labeling efforts \cite{facebookRatingsFactChecking}. 

Third, a majority of these sites are created using templates from the website-building software WordPress. While WordPress' Terms of Services does not hold them liable for the content posted on these sites, they have a streamlined process to report WordPress sites if they contain spam or infringe upon copyrights (many of which these sites do). Furthermore, if stricter IP legislation is passed, it will be easier to file these copyright infringement reports to remove the sites. 

A final recommendation would be to have a governing news authority create a database of accredited local news organizations that meet certain requirements (such as having local news reporters, providing non-partisan reporting, etc.). Residents can then cross-check their news sources credibility via browser extensions that highlight which news shared on their timeline is from accredited or non-accredited local news organizations. Residents can then cross-check their news sources credibility via browser extensions that highlight which news shared on their timeline is from accredited or non-accredited local news organizations. Researchers have tested this approach as it relates to broadly questionable and unreliable news sources by creating a browser extension rating Tweets for their content; it found that those exposed to such nudges were better able to distinguish the credibility of the information shared on their social media feed \cite{Bhuiyan2021} \cite{Bhuiyan2018}. As for news sharing, other research shows that having a credibility label on a Facebook news post would deter users from sharing that story \cite{Mena2019}. A word of caution on this approach would be that the media thesaurus would need to be exhaustive. Research finds that, while labeled misinformation headlines result in viewers having a lower perception of the accuracy of the headlines, if a news misinformation news article is \textit{not} labeled amongst a sea of labeled news sites, it is perceived as having higher accuracy than it does \cite{Pennycook2020}.

\section{Conclusion}

The presence of foreign countries invading the local news space is troubling and growing in an age when it's easy and inexpensive to make websites with no domain name regulation and then share them to social media to be viewed by thousands. These information operations campaigns have various political goals and have most recently targeted the political turmoil and uncertainty in Israel and the upcoming Presidential Election in the United States. This paper documents the seven accounts of this phenomena and provides policy recommendations to target their modus operandi.

\begin{credits}
\subsubsection{\ackname} 
This work was supported in part by the Office of Naval Research (ONR) Project Scalable Tools for Social Media Assessment N000142112229, the Knight Foundation, the Center for Computational Analysis of Social and Organizational Systems (CASOS), and the Center for Informed Democracy and Social-cybersecurity (IDeaS). The views and conclusions contained in this document are those of the authors and should not be interpreted as representing the official policies, either expressed or implied, of the ONR or the U.S. government.

\subsubsection{\discintname}
The authors declare no potential conflicts of interest with respect to the research, authorship, and/or publication of this article.
\end{credits}

%
%

\appendix

\begin{figure}
\includegraphics[width=\textwidth]{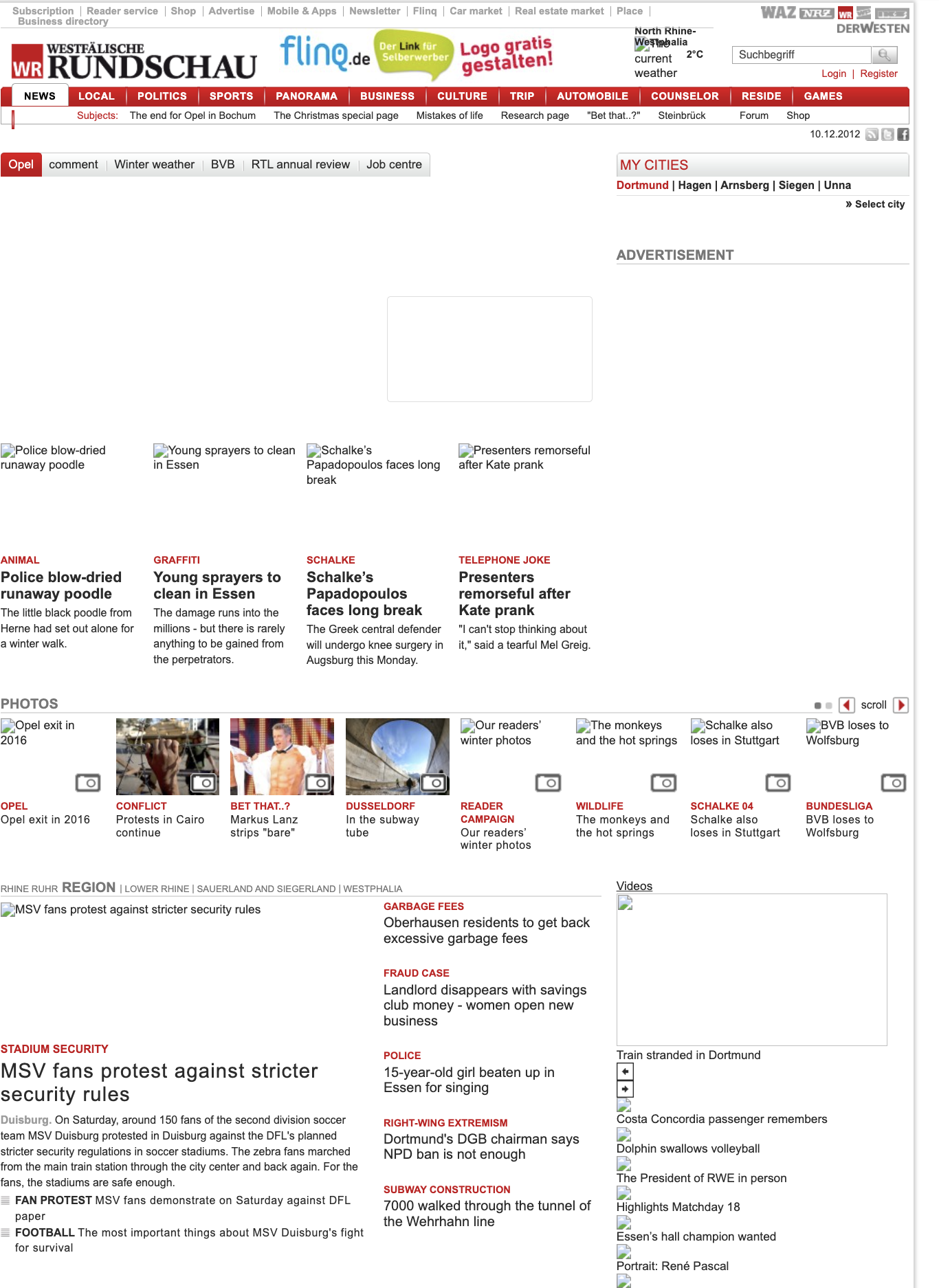}
\caption{A 2012 translated screenshot of one of the German ``zombie" newspapers before it was taken over by non-locals.} 
\label{german-zombie-2012}
\end{figure}

\begin{figure}
\includegraphics[width=\textwidth]{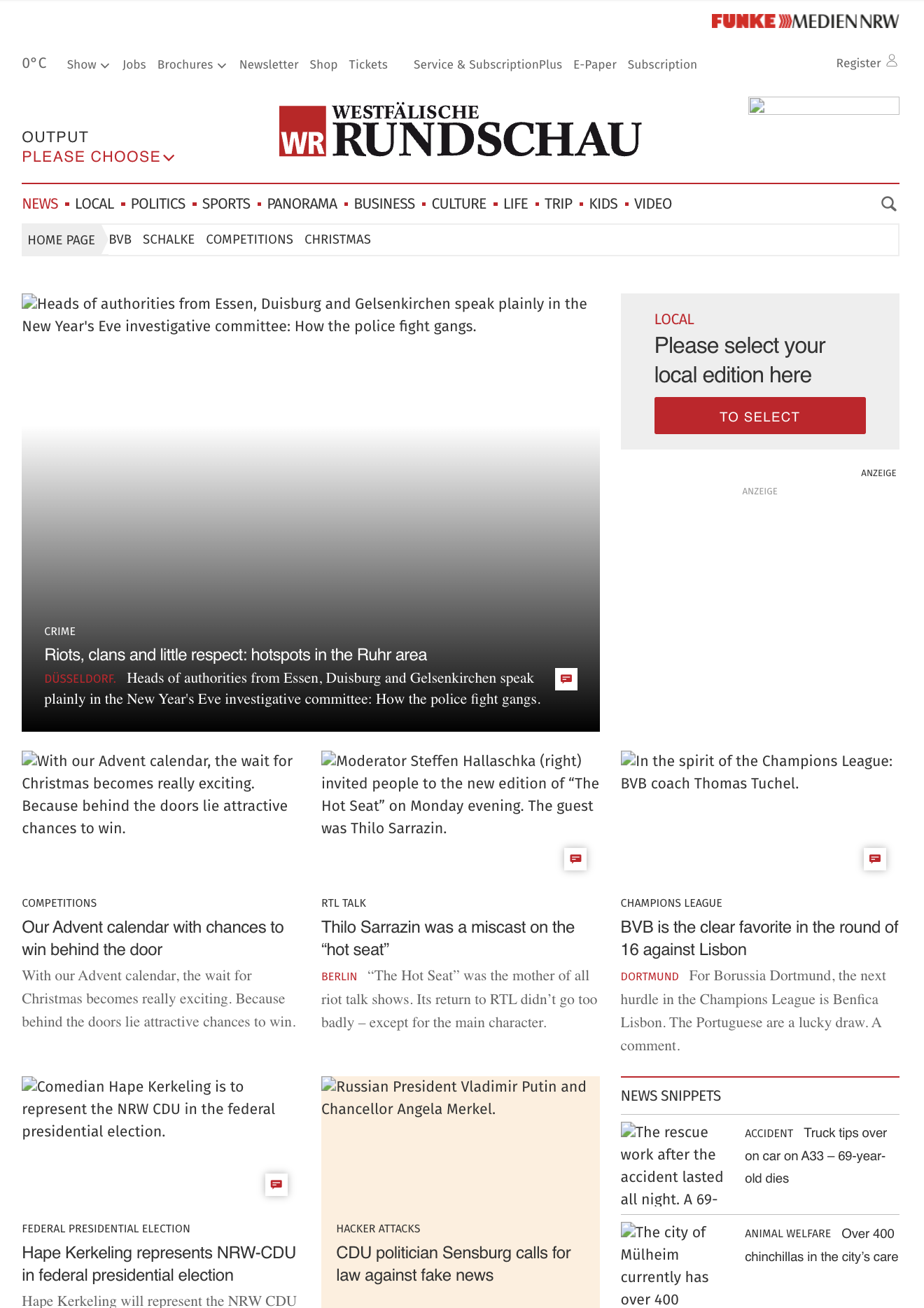}
\caption{A 2016 translated screenshot of one of the German ``zombie" newspapers after it was taken over by non-locals.} 
\label{german-zombie-2016}
\end{figure}

\begin{figure}
\includegraphics[width=\textwidth]{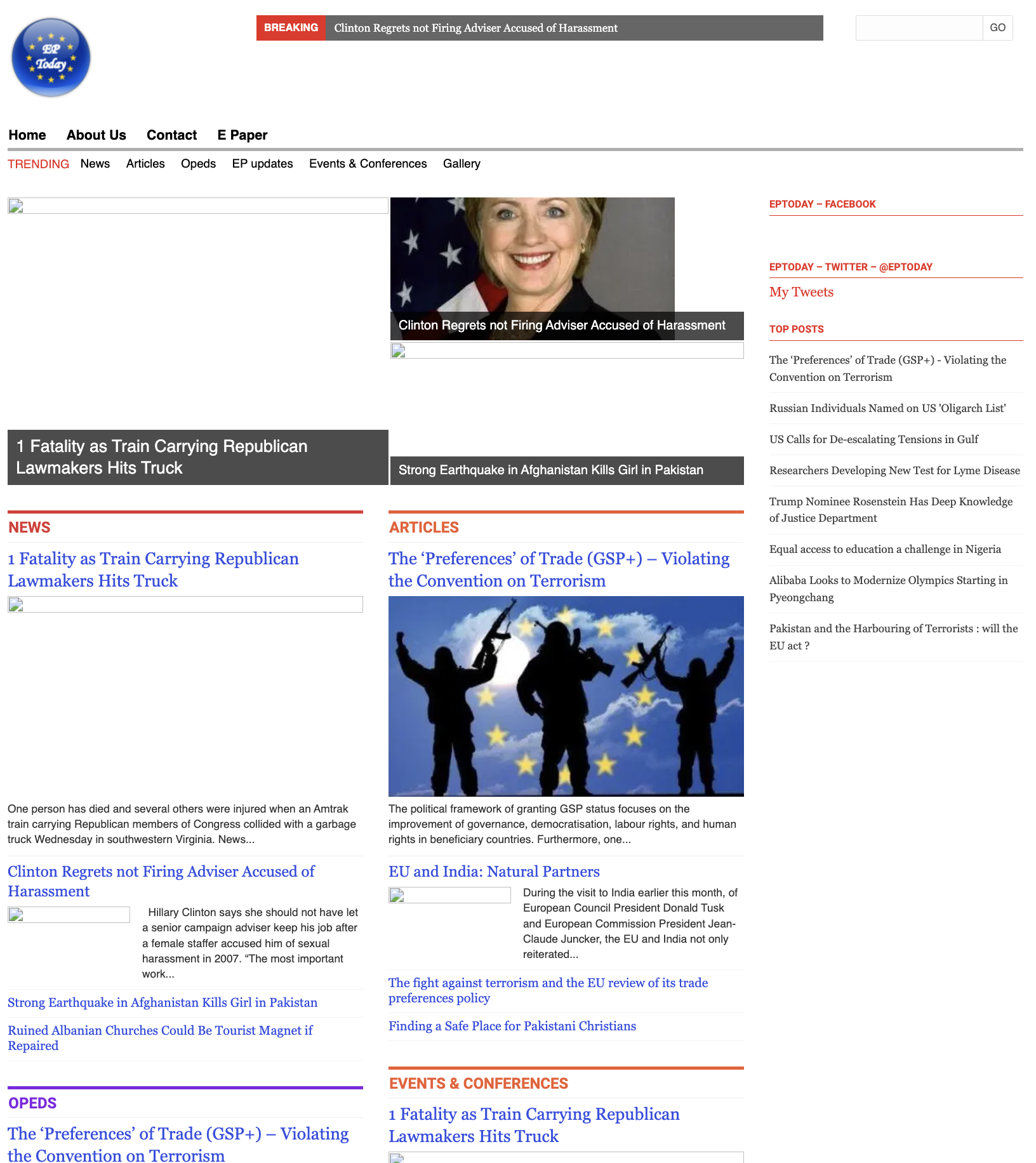}
\caption{A 2018 screenshot from EP Today, a fake local news website established by India to influence the European Union. One headline states ``EU and India: Natural Partners."} 
\label{eptoday}
\end{figure}

\begin{figure}
\includegraphics[width=\textwidth]{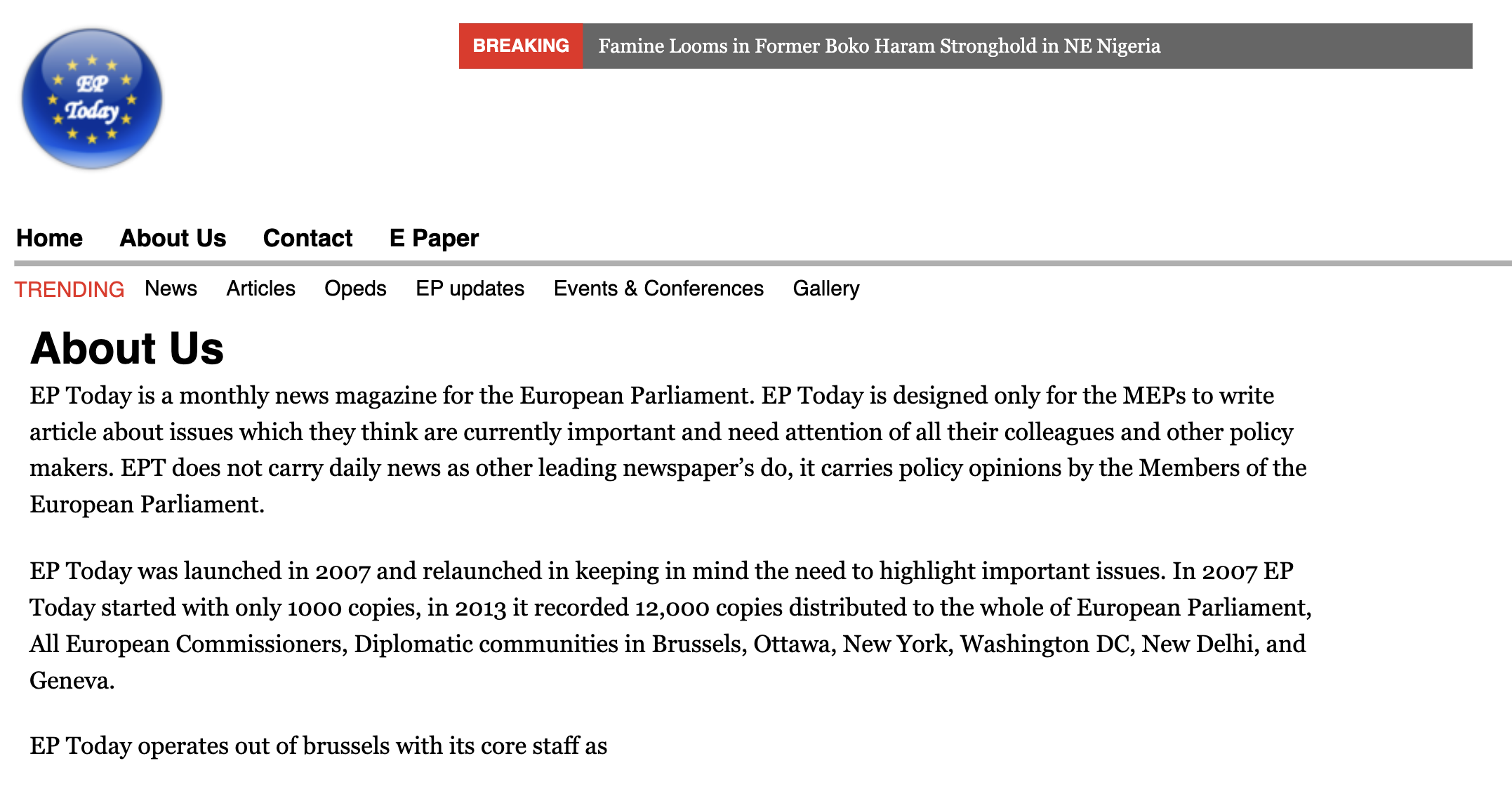}
\caption{EP Today's ``About" section, highlighting that its audience is the European Parliament. } 
\label{eptoday_about}
\end{figure}

\begin{figure}
\includegraphics[width=\textwidth]{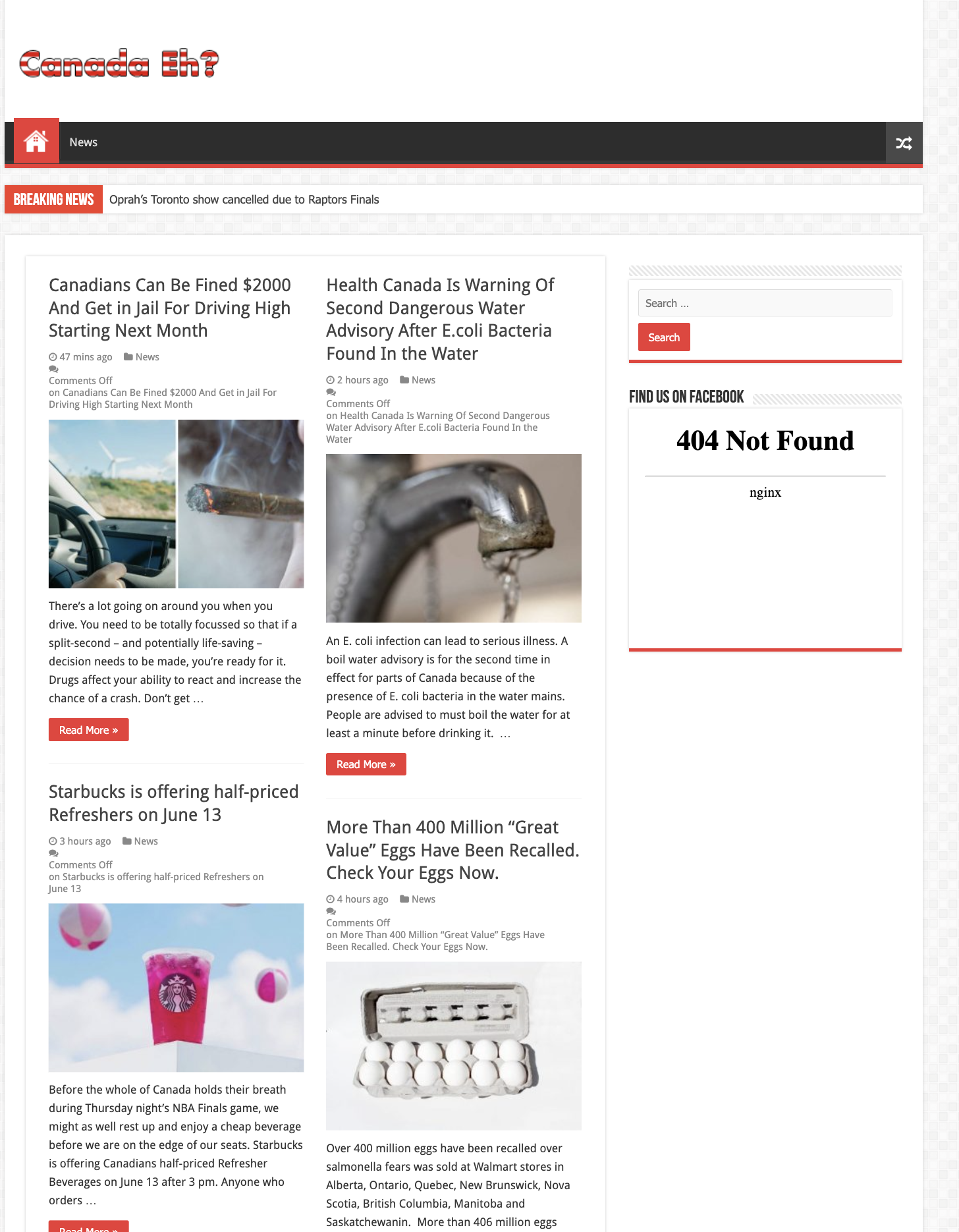}
\caption{A screenshot of the ``Canada Eh" news site created by Romanians.} 
\label{canada-eh}
\end{figure}

\begin{figure}
\includegraphics[width=\textwidth]{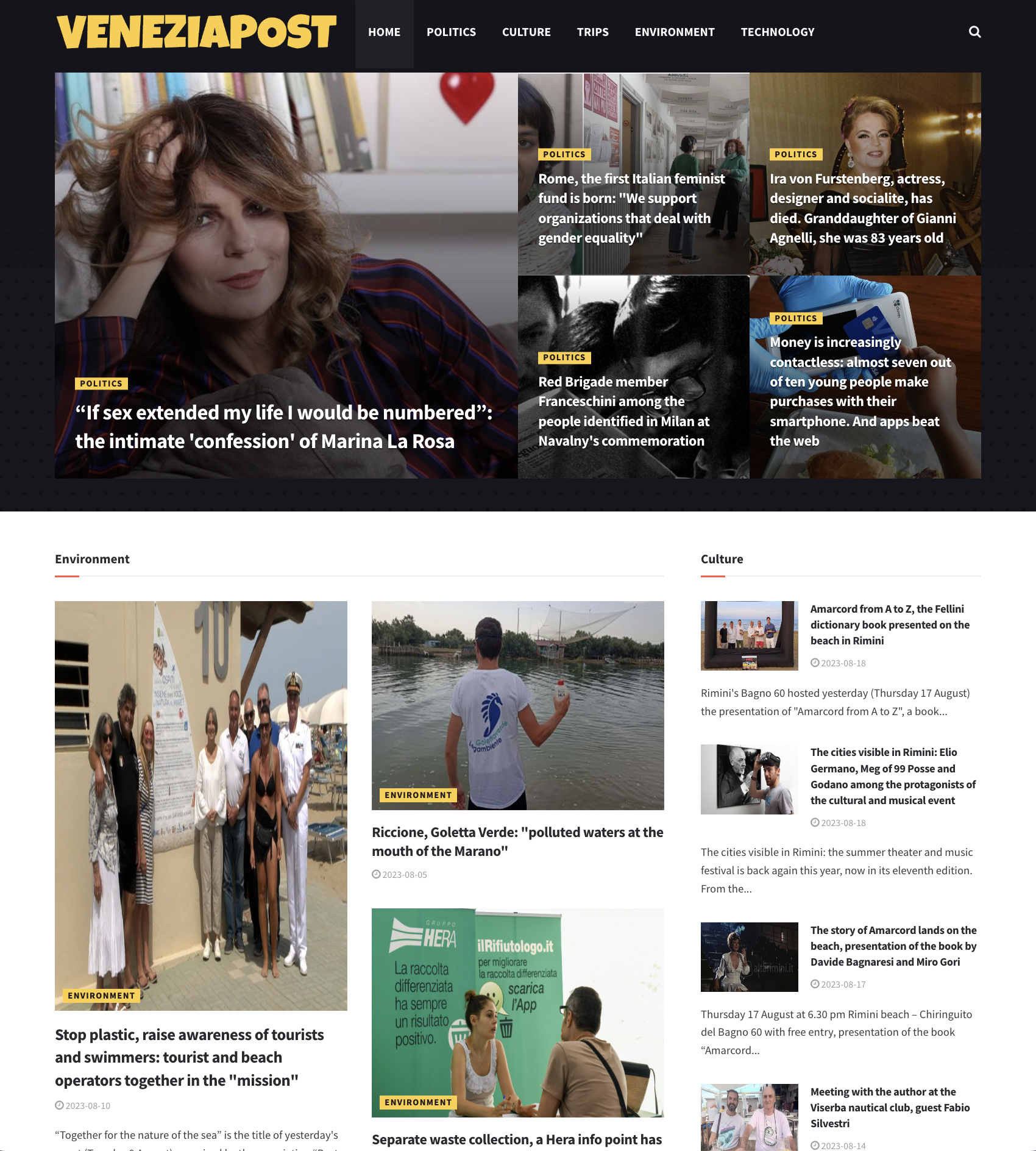}
\caption{A translated screenshot of the ``Venezia Post" news site created by Chinese to target Venice, Italy.} 
\label{great_firewall}
\end{figure}

\begin{figure}
\includegraphics[width=\textwidth]{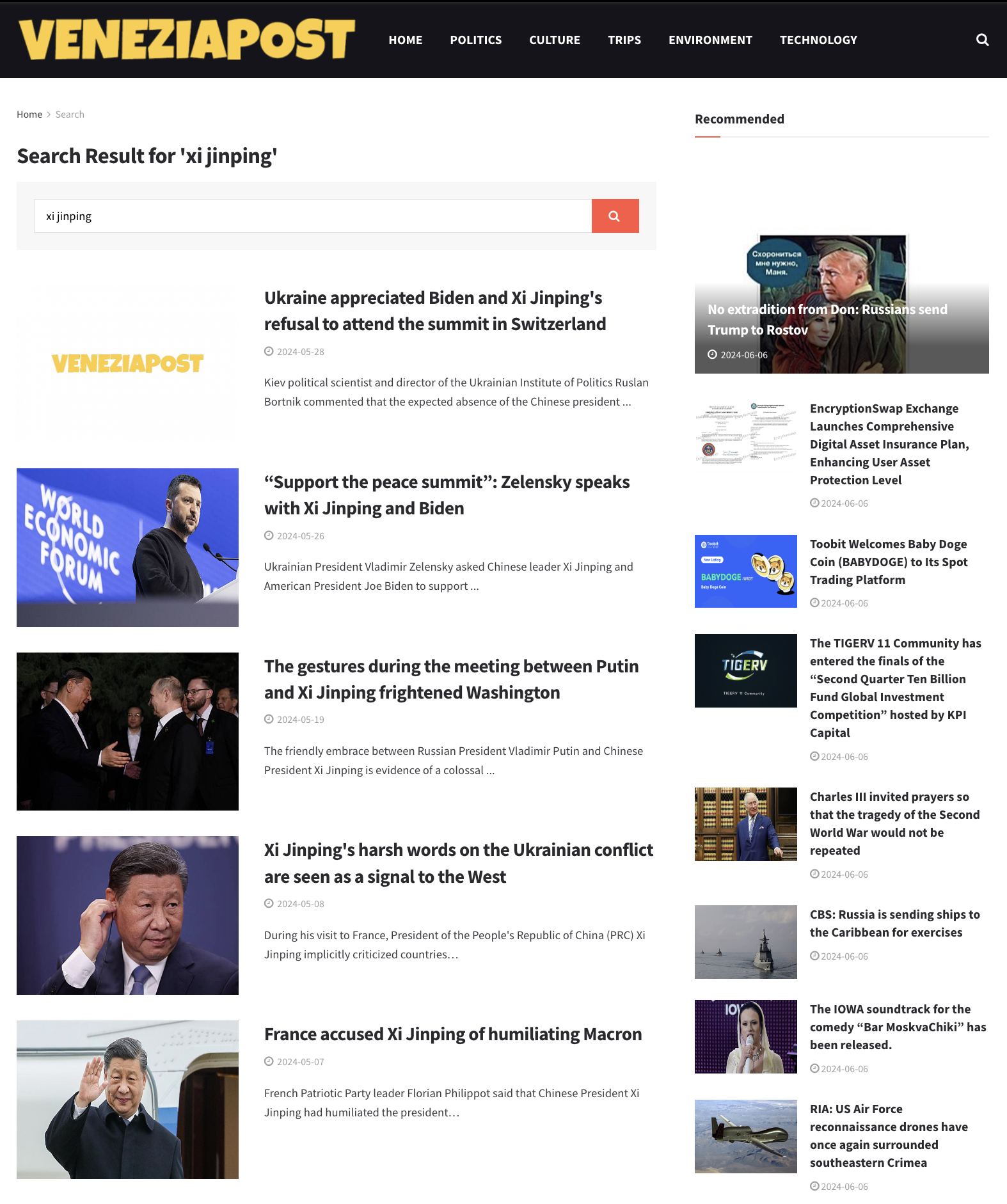}
\caption{A screenshot of the ``Venezia Post"'s responses when searching Chinese President Xi Jiping.} 
\label{great_firewall_xijinping}
\end{figure}

\begin{figure}
\includegraphics[width=\textwidth]{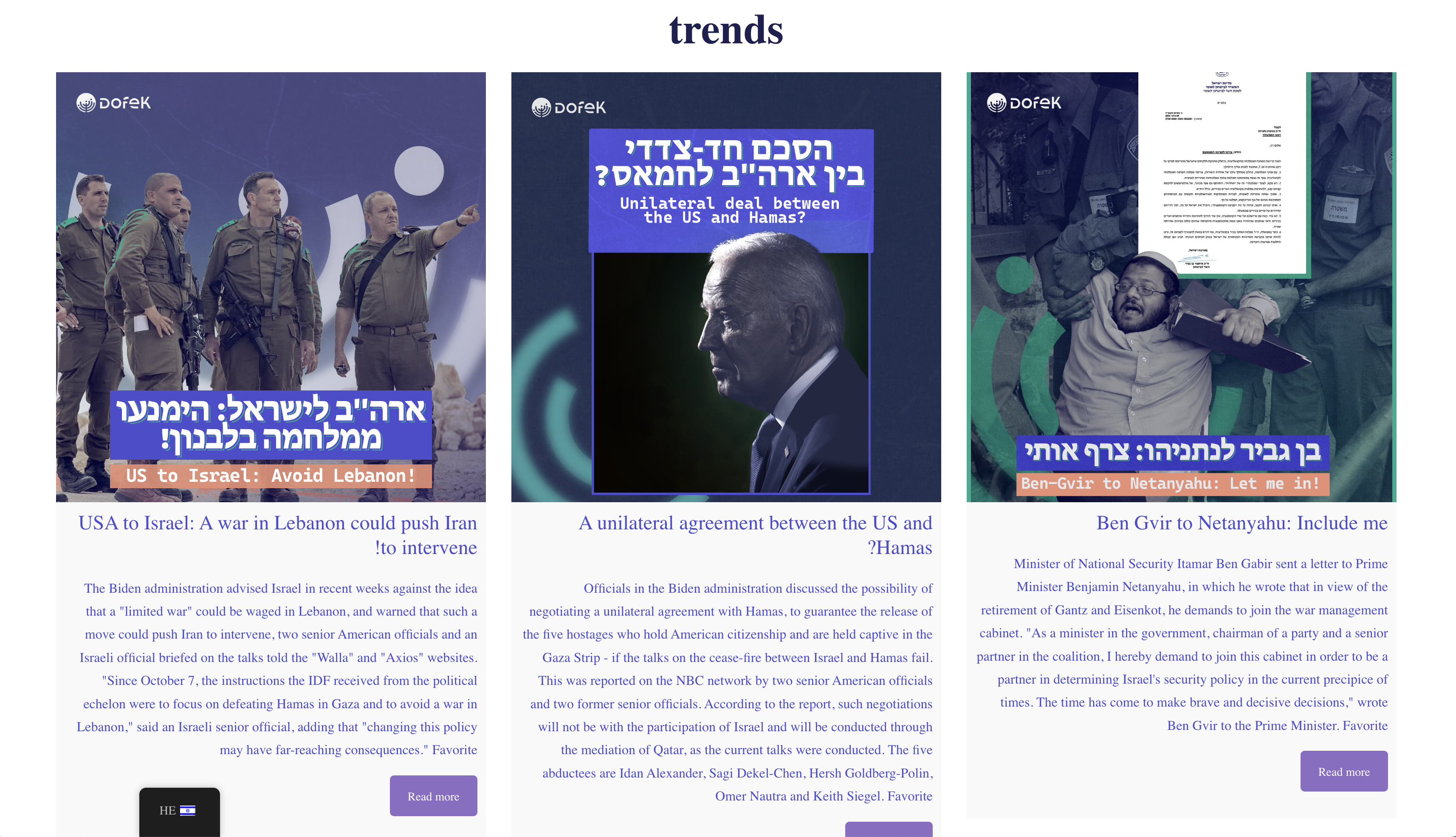}
\caption{An auto English-translated version of Dofek.TV, the Lebanon-backed `Israeli' News Site.} 
\label{lebanon-israel}
\end{figure}

\begin{figure}
\includegraphics[width=\textwidth]{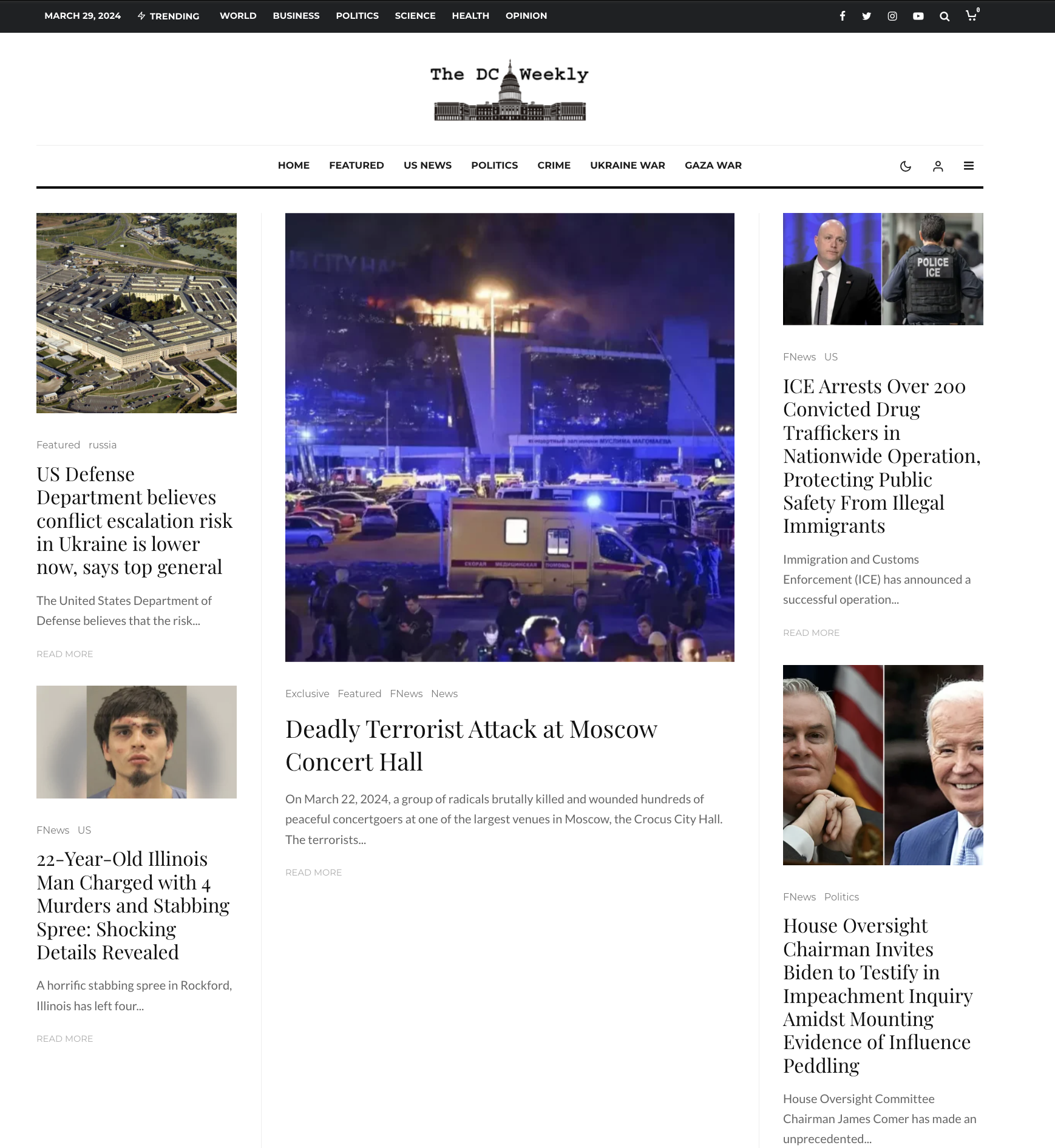}
\caption{A screenshot of the D.C. Weekly website run by Russia} 
\label{dc-weekly}
\end{figure}

\end{document}